# Interwell Simulation Model[1] for the Advection Dispersion Equation (ISADE)

Mohammad S. Jamal[1], Abeeb A. Awotunde[1], Mohammed S. Al-Kobaisi[2], Hasan Y. Al-Yousef[1], Ahmed Sadeed[1] and Shirish Patil[1]

[1]College of Petroleum Engineering & Geosciences, King Fahd University of Petroleum & Minerals, Dhahran, K.S.A.
[2]Department of Petroleum Engineering, Khalifa University of Science and Technology, Abu Dhabi, U.A.E.

**Abstract**

Contaminant transport in porous media is often modeled by solving the Advection Dispersion Equation (ADE) which describes the way the contaminant moves within the bulk fluid as well as the movement of the contaminant due to the bulk movement of the fluid phase. The simulation of contaminant transport can however prove to be computationally expensive especially if the porous medium is discretized into high-resolution grid blocks.

We propose an interwell simulation model for advection dispersion equations (ISADE) to predict the concentration of contaminant observed at the pumping well. This method comprises of two main steps. Initially, the model divides the aquifer or reservoir into a series of 1D injector-producer pairs and uses the historical contaminant observation data to estimate five major unknowns in each of these control volumes: the interwell connectivity, the pore volume, the volumetric flowrate at the grid face, the dispersion coefficient, and the number of grids in each control volume. Finally, once the history matching process is complete, the estimated variables are used to predict the concentration of contaminant observed at the wells. Five example cases were studied and the results show that the ISADE model was able to predict with reasonable accuracy the concentration of contaminant observed. Furthermore, the model can easily handle changes in input parameters such as the concentration of contaminants released in the aquifer, and the injection and production rates.

## 1. Introduction

Advection Dispersion Equation (ADE) is widely implemented to model several different transport phenomena occurring in porous media. One of the major applications of ADE in groundwater aquifers is modeling of the transport of contaminants such as salts. The intrusion of seawater into coastal aquifers increases the salinity of groundwater and therefore makes it non-potable. It is essential that the transport of salt in the aquifer be modeled with reasonable accuracy to sustainably plan water production from these coastal aquifers (Arfib & Charlier, 2016; Benson et al., 1998; Dokou & Karatzas, 2012; Kourakos & Mantoglou, 2009; Mongelli et al., 2013; Sherif et al., 1988; Sreekanth & Datta, 2015). Another application of ADE in porous media is the modeling of tracer transport. Tracer tests are generally used for the exploration and characterization of aquifers and reservoirs. These tests are an important tool in understanding the flow paths of groundwater in underground aquifers and are used to determine the direction and velocity of flow, as well as the presence of highly permeable channels (Maliva, 2016; Pruess, 2002; Wei et al., 1990; Zheng & Jiao, 1998). Single Well Chemical Tracer Tests (SWCTT) are also widely employed in the petroleum industry to estimate the remaining oil saturation after water-flooding (Huseby et al., 2012; Jerauld et al., 2010). The ADE is also used to model transport of chemicals such as surfactants, nanoparticles and polymers in porous media to enhance the recovery of oil from reservoirs (Chung, 1991; Gao et al., 1993; Illiano et al., 2021; Manzoor, 2020). Some of these chemicals are expensive and it is therefore important to predict the effects of these chemicals on oil production before implementing it in the field.

The Advection Dispersion Equation is derived by substituting the dispersive flux (approximated by the Fick's first law) and the advective flux into an appropriately defined continuity equation. The advective transport refers to the transport of contaminants by the bulk flow of fluid and requires that the velocity of the fluid is known. The dispersion term represents the spread of contaminants within the fluid (and in some cases within the solid rock) by different microscopic and macroscopic effects. The macroscopic dispersion is also dependent on the velocity of the fluid and is therefore directly related to the physical properties (i.e. permeability and porosity) of the aquifer (Bear, 1979; Sharma et al., 2020). The microscopic diffusion or molecular diffusion occurs when the contaminants move from a higher to a lower concentration. The finite-difference scheme is sometimes employed to solve the governing partial differential equation defining the transport of contaminants in a porous medium (Ataie-Ashtiani & Hosseini, 2005; Bear & Cheng, 2010; Chakraborty & Ghosh, 2013; Jamal & Awotunde, 2018; Jamal & Awotunde, 2020; Liu et al., 2011; Sun, 1989). Finite-difference scheme solves the governing partial differential equation by converting it into a system of discrete equations. For a single contaminant flowing

---

in a single fluid phase, the number of discrete equations equals the number of grids used in discretizing the spatial system. The computational expense associated with solving this system of equations increases as the number of grids increases. The system of equations developed is solved for each time-step. The process of contaminant transport could be slow and may occur over a long period of time (i.e., could occur over several years) (Fiori et al., 2019). Therefore, to obtain reliable results the entire time period is discretized into a large number of smaller time-steps. This further increases the computational overhead. Similar problems exists when simulating fluid flow on a full-scale reservoir using conventional grid systems. Additionally, modeling flow in reservoirs becomes even more complicated due to the involvement of geological heterogeneities, such as permeability and porosity. Although several attempts have been made to improve the computational efficiency of the simulation of contaminant transport, e.g. MT3D (Zheng and Papadopulos 1990), these codes still often solve the problem in three-dimensional domain thus incurring considerable computational costs. In addition, several proxy models have been developed to predict the production rate while reducing the computational overhead. One commonly used proxy model is the capacitance resistance model (CRM) (Sayarpour et al., 2009; Yousef et al., 2006). The CRM is a semi-theoretical model that neglects flux term in the fluid continuity equation. The model relates fluid rates at the pumping wells to fluid rates and pressures at the injection wells without considering the fluid fluxes in the reservoir. This makes the model runs fast but subject to inaccuracies due to the neglected terms in the continuity equation. The primary unkonwns for each injector-producer in the CRM method are the time-constant, the interwell conneccitivity and the productivity index. The interwell connectivity factor between an injector $i$ and a particular producer $j$ is the fraction of the injector's energy directed towards that producer (Sayarpour, 2008). Its value can vary from zero to one. A value of zero indicates that there is no contribution from the injector towards a producer. These unknowns are estimated by history matching the known production data. Another commonly used method is the Interwell Numerical Simulation Model (INSIM) where the reservoir is characterized as a series of 1D connections between well pairs with the unkown parameters for each pair being the transmissibility and the pore volume (Guo et al., 2018; Zhao et al., 2015, 2016).

These proxy models are computationally cheap as they require few parameters to predict the production rates at the wells. Currently there are no known interwell models to predict contaminant concentrations at the pumping wells. Naudomsup & Lake, 2019, used a combination of analytical dispersion-only model (Ogata & Banks, 1961), and the Koval model (Koval, 1963), to estimate the reservoir characteristics by matching the tracer concentration data. In this paper, an interwell simulation model for advection dispersion equation (ISADE) is developed for predicting contaminant/tracer concentrations observed at pumping wells. ISADE is a semi-theoretical model that uses a series of one-dimensional connections between each producer-injector pair with all pairs connected by interconnectivity factors similar to those found in CRM models. The ADE is then solved for each of these control volumes separately. One advantage of the ISADE model is that only the flowrates at the wells are required to run the model. In case of the full-scale simulation model velocities at the interfaces of grid blocks are required to solve the ADE.

Five examples were presented to test the effectiveness of the ISADE model. The first three examples each involve a synthetic homogeneous aquifer with different well configurations. The other two examples considered are more realistic heterogenous aquifers with varying concentrations of the contaminant released into the aquifers at the injection wells. All the examples studied show that the results obtained from ISADE closely match those from the full-scale simulation model.

## 2. The Advection Dispersion Equation

The advection dispersion equation is given as

$$\nabla \cdot (\phi \tilde{D} \cdot \nabla c) - \nabla \cdot (\vec{u} c) + \frac{I c_{inj}}{V_b} - \frac{Q c_{prd}}{V_b} = \frac{\partial (\phi c)}{\partial t}, \qquad (1)$$

where, $\phi$ is the porosity, $\tilde{D}$ is the dispersion coefficient tensor $(L^2 T^{-1})$, $\vec{u}$ is the velocity vector of the fluid $(LT^{-1})$, $c$ is the concentration of the tracer $(ML^{-3})$, $c_{inj}$ is the concentration of the contaminant released into the aquifer $(ML^{-3})$, $c_{prd}$ is the concentration of the contaminant removed from the aquifer $(ML^{-3})$, $V_b$ is the block volume $(L^3)$, $I$ is the volumetric flow rate at the injection wells $(L^3 T^{-1})$, and $Q$ is the volumetric flow rate at the production wells $(L^3 T^{-1})$. Note that both the production and injection rates have positive values.

The dispersion coefficient tensor when the axis is aligned to the velocity directions is given by $\tilde{D} = \begin{pmatrix} D_x & 0 \\ 0 & D_y \end{pmatrix}$, where, $D_x$ and $D_y$, are the principal terms of the dispersion coefficient tensor in the $x$- and $y$- directions, respectively. The equations for the dispersion coefficients are given as, (Zhang & Bennett, 2002)

$$D_x = \alpha_L \frac{u_x^2}{|u|} + \alpha_T \frac{u_y^2}{|u|} + D^*, \qquad (2)$$

and,

$$D_y = \alpha_L \frac{u_y^2}{|u|} + \alpha_T \frac{u_x^2}{|u|} + D^*, \qquad (3)$$

where, $D^*$, is the molecular diffusion coefficient, $u_x$ is the velocity of fluid flow in *x*-direction and $u_y$ is the velocity of fluid flow in the *y*-direction, and,

$$|u| = \sqrt{u_x^2 + u_y^2}. \qquad (4)$$

$\alpha_L$ and $\alpha_T$ are the longitudinal and transverse dispersivity.

## 3. Interwell Simulation of Advection Dispersion Equation (ISADE)

Figure 1 displays a simple two-dimensional model of an aquifer which has been spatially discretized into $H \times L$ grids, where, $H$ is the total number of grids in the *x*-direction and $L$ is the total number of grids in the *y*-direction. There are four injectors (one at each corner of the reservoir) and a single producer at the centre. The interest here is to determine the concentration of the contaminant observed at the pumping well. The contaminant transport in the reservoir/aquifer is modelled using the advection dispersion equation which is solved by discretizing the partial differential equation given by Eq. 1. The discretized form of Eq. 1 for a two-dimensional system (gridded in two directions *x* and *y*) is given as

$$
\begin{aligned}
&\left(\frac{\phi D_y}{\Delta y^2}\right)^{n+1}_{h,l-\frac{1}{2}} c^{n+1}_{h,l-1} + \left(\frac{\phi D_x}{\Delta x^2}\right)^{n+1}_{h-\frac{1}{2},l} c^{n+1}_{h-1,l} \\
&- \left[\left(\frac{\phi D_x}{\Delta x^2}\right)^{n+1}_{h+\frac{1}{2},l} + \left(\frac{\phi D_x}{\Delta x^2}\right)^{n+1}_{h-\frac{1}{2},l} + \left(\frac{\phi D_y}{\Delta y^2}\right)^{n+1}_{h,l+\frac{1}{2}} + \left(\frac{\phi D_y}{\Delta y^2}\right)^{n+1}_{h,l-\frac{1}{2}} + \frac{\phi_{h,l}}{\Delta t}\right] c^{n+1}_{h,l} \\
&+ \left(\frac{\phi D_x}{\Delta x^2}\right)^{n+1}_{h+\frac{1}{2},l} c^{n+1}_{h+1,l} + \left(\frac{\phi D_y}{\Delta y^2}\right)^{n+1}_{h,l+\frac{1}{2}} c^{n+1}_{h,l+1} \\
&- \frac{u^{n+1}_{x_{h+\frac{1}{2},l}}}{\Delta x} c^{n+1}_{h+\frac{1}{2},l} + \frac{u^{n+1}_{x_{h-\frac{1}{2},l}}}{\Delta x} c^{n+1}_{h-\frac{1}{2},l} - \frac{u^{n+1}_{y_{h,l+\frac{1}{2}}}}{\Delta y} c^{n+1}_{h,l+\frac{1}{2}} + \frac{u^{n+1}_{y_{h,l-\frac{1}{2}}}}{\Delta y} c^{n+1}_{h,l-\frac{1}{2}} = -\frac{\phi_{h,l}}{\Delta t} c^n_{h,l} - \frac{Ic_{inj}}{V_{b_{h,l}}} + \frac{Qc_{prd}}{V_{b_{h,l}}}
\end{aligned}
\qquad (5)
$$

where, $h$ and $l$ are the indices of the grid blocks in the *x* and *y*-directions, respectively, $\Delta x$ and $\Delta y$ are the grid dimensions in the *x*- and *y*-directions, respectively and $n$ is the index of time. The discretization is done such that the concentration is calculated at the grid centers and the velocities are given at the boundaries of each grid. The velocities at the grid interfaces are calculated from the pressures obtained from solving the fluid flow equation. For the advective part, the concentrations at the interfaces are calculated by an upwinding technique. For the example shown in Fig. 1, a system of $M = H \times I$ simultaneous equations results from Eq. 5. If the original PDE (Eq. 1) is linear, then the system of equations resulting from Eq. 5 will also be linear and can be written as $A\vec{x} = \vec{b}$, where $A$ is a coefficient matrix of the size $M \times M$, $\vec{x}$ is a solution vector of size $M \times 1$ and $\vec{b}$ is a vector of size $M \times 1$, and $M$ is the total number of grid blocks. The computational expense in solving the set of linear equations increases as the number of grid blocks increases.

The ISADE model, however, allows for the partitioning of the reservoir into smaller control volumes, each consisting of a unique injector-producer pair. Figure 2 shows the partitioning of the example discussed in Fig. 1 into four control volumes. Each of these control volumes is assumed to be one dimensional as shown in Fig. 3 with an injector and producer at each end. Thus, ISADE uses the same discretized equation from the full-scale simulation (Eq. 5) albeit in only one-dimension connecting a producer to an injector. Equation 5 for a 1D case can be written as

$$\left(\frac{\phi D_x}{\Delta x^2}\right)^{n+1}_{k-\frac{1}{2}} c^{n+1}_{k-1} - \left[\left(\frac{\phi D_x}{\Delta x^2}\right)^{n+1}_{k+\frac{1}{2}} + \left(\frac{\phi D_x}{\Delta x^2}\right)^{n+1}_{k-\frac{1}{2}} + \frac{\phi_k}{\Delta t}\right] c^{n+1}_k + \left(\frac{\phi D_x}{\Delta x^2}\right)^{n+1}_{k+\frac{1}{2}} c^{n+1}_{k+1},$$

$$-\frac{u^{n+1}_{x_{k+\frac{1}{2}}}}{\Delta x} c^{n+1}_{k+\frac{1}{2}} + \frac{u^{n+1}_{x_{k-\frac{1}{2}}}}{\Delta x} c^{n+1}_{k-\frac{1}{2}} = -\frac{\phi_k}{\Delta t} c^n_k - \frac{Ic_{inj}}{V_{b_k}} + \frac{Qc_{prd}}{V_{b_k}}$$

(6)

where, $k$ is the index of the gridblock in the smaller control volume.

Since there are typically several producers and injectors in a reservoir, ISADE describes the contaminant transport within each injector-producer pair (i.e., each control volume) using the one-dimensional discretized equation in Eq. 6 and then connects all the equations from all the injector-producer pair with appropriately defined interwell connectivity factors $\left(f_{i,j}\right)$.

The ISADE model assumes that within any specific control volume,

i) The porosity and the block volumes are constant with respect to space and time.
ii) The dispersion coefficient is homogeneous and the velocity field is spatially uniform. However, these two parameters can change with time.

With these assumptions, Eq. 6 for each control volume reduces to

$$\frac{\phi_{i,j} V_{b_{i,j}} D^{n+1}_{x_{i,j}}}{\Delta x^2_{i,j}} \left(c^{n+1}_{k_{i,j}-1} - 2c^{n+1}_{k_{i,j}} + c^{n+1}_{k_{i,j}+1}\right) - \frac{\phi_{i,j} V_{b_{i,j}}}{\Delta t} c^{n+1}_{k_{i,j}} - \frac{V_{b_{i,j}} u^{n+1}_{x_{i,j}}}{\Delta x_{i,j}} \left(c^{n+1}_{k_{i,j}+\frac{1}{2}} - c^{n+1}_{k_{i,j}-\frac{1}{2}}\right),$$

$$= -\frac{\phi_{i,j} V_b}{\Delta t} c^n_{k_{i,j}} - I^{n+1}_{i,j} c^{n+1}_{inj_{i,j}} + Q^{n+1}_{i,j} c^{n+1}_{prd_{i,j}}$$

(7)

where, $i$ is the index for injectors, $j$ is the index for producers, $(i, j)$ in the subscript denotes the value of the parameter for the control volume connecting injector $i$ to producer $j$. Substituting $V_{p_{i,j}} = \phi_{i,j} V_{b_{i,j}}$ and $q^{n+1}_{f_{i,j}} = \frac{V_{b_{i,j}} u^{n+1}_{x_{i,j}}}{\Delta x_{i,j}}$ in Eq. 7 gives

$$\frac{V_{p_{i,j}} D^{n+1}_{x_{i,j}}}{\Delta x^2_{i,j}} \left(c^{n+1}_{k_{i,j}-1} - 2c^{n+1}_{k_{i,j}} + c^{n+1}_{k_{i,j}+1}\right) - \frac{V_{p_{i,j}}}{\Delta t} c^{n+1}_{k_{i,j}} - q^{n+1}_{f_{i,j}} \left(c^{n+1}_{k_{i,j}+\frac{1}{2}} - c^{n+1}_{k_{i,j}-\frac{1}{2}}\right).$$

$$= -\frac{V_{p_{i,j}}}{\Delta t} c^n_{k_{i,j}} - I^{n+1}_{i,j} c^{n+1}_{inj_{i,j}} + Q^{n+1}_{i,j} c^{n+1}_{prd_{i,j}}$$

(8)

In cases when there is only a single injector-producer pair, $I^{n+1}_{i,j} c^{n+1}_{inj_{i,j}} = I^{n+1}_i c^{n+1}_{inj_i}$ and $Q^{n+1}_{i,j} c^{n+1}_{prd_{i,j}} = Q^{n+1}_j c^{n+1}_{prd_j}$. However, in cases where more than one injector-producer pair exist, the concentration of contaminant observed at a pumping well comes from several injection wells. Under such conditions, the well interwell connectivity factor, $f_{i,j}$, is introduced to account for the fraction of the contaminat at the pumping well that is attributable to each injector well. This is similar to the concept of interwell connecitivity used in CRM models (Sayarpour et al., 2009; Yousef et al., 2006). Thus, for the cases of multiple wells in the reservoir, the source term is rewritten as

$$I^{n+1}_{i,j} c^{n+1}_{inj_{i,j}} = f_{i,j} I^{n+1}_i c^{n+1}_{inj_i}.$$

(9)

The interwell connectivity factors are constrained by

$$\sum_{j=1}^{n_p} f_{i,j} \leq 1,$$

(10)

where, $n_p$ is the total number of producers.

To solve Eq. 8, the flow rate $Q_{i,j}$ in any well $j$ resulting from its interaction with injector $i$, is required to be known. This can be computed from

$$Q_{i,j}^{n+1} = \frac{f_{i,j} I_i^{n+1}}{\sum_{i=1}^{n_{inj}} f_{i,j} I_i^{n+1}} Q_j^{n+1} \tag{11}$$

where $n_{inj}$ is the total number of injectors and the total rate $Q_j$ in producer $j$ is known *a priori* from the fluid flow model. Also, in Eq. 8, $c_{prd_{i,j}}$ is the concentration of contaminant observed at pumping well $j$ due to the release of contaminant at injector $i$. The total contaminant concentration observed at each producer can then be calculated as

$$c_{prd_j}^{n+1} = \sum_{i=1}^{n_{inj}} c_{prd_{i,j}}^{n+1}. \tag{12}$$

The grid block dimension $\Delta x_{i,j}$ for each control volume can be calculated as follows

$$\Delta x_{i,j} = \frac{L_{well_{i,j}}}{N_{div_{i,j}}}, \tag{13}$$

where, $L_{well_{i,j}}$ is the length between injector $i$ and producer $j$, and $N_{div_{i,j}}$ is the number of division each control volume is discretized into. Substituting Eqs. 9, 11 and 13 into Eq. 8, we obtain

$$\frac{V_{p_{i,j}} D_{x_{i,j}}^{n+1} N_{div_{i,j}}^2}{L_{well_{i,j}}^2} \left( c_{k_{i,j}-1}^{n+1} - 2 c_{k_{i,j}}^{n+1} + c_{k_{i,j}+1}^{n+1} \right) - \frac{V_{p_{i,j}}}{\Delta t} c_{k_{i,j}}^{n+1} - q_{f_{i,j}}^{n+1} \left( c_{k_{i,j}+\frac{1}{2}}^{n+1} - c_{k_{i,j}-\frac{1}{2}}^{n+1} \right)$$
$$= -\frac{V_{p_{i,j}}}{\Delta t} c_{k_{i,j}}^n - f_{i,j} I_i^{n+1} c_{inj_i}^{n+1} + \frac{f_{i,j} I_i^{n+1}}{\sum_{i=1}^{n_{inj}} f_{i,j} I_i^{n+1}} Q_j^{n+1} c_{prd_{i,j}}^{n+1} \tag{14}$$

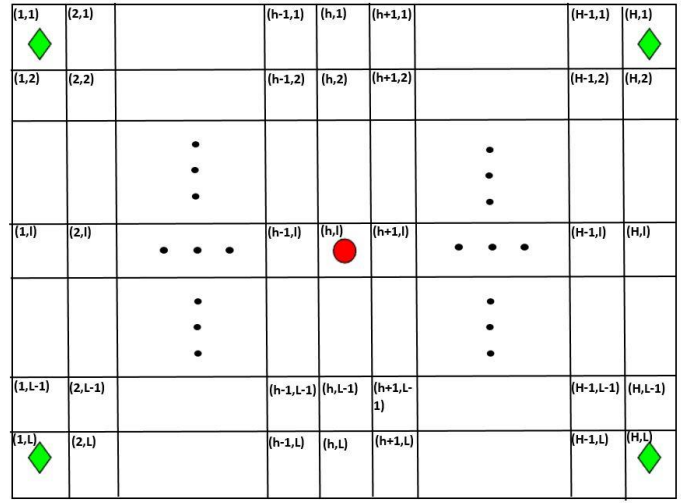

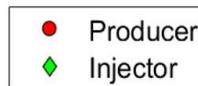

**Fig. 1** Figure displaying the spatial discretization of a simple two-dimensional aquifer model consisting of four injectors and one producer.

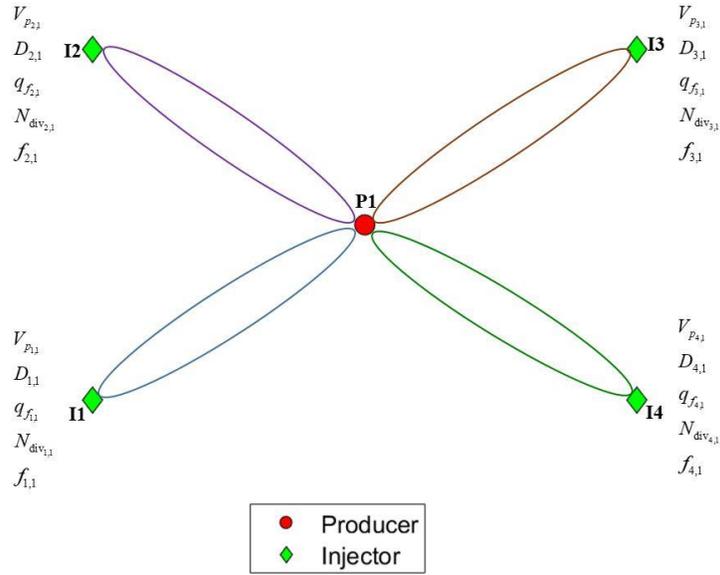

**Fig. 2** Division of reservoir shown in Fig. 1 into four separate control volumes, each consisting of an injector-producer pair.

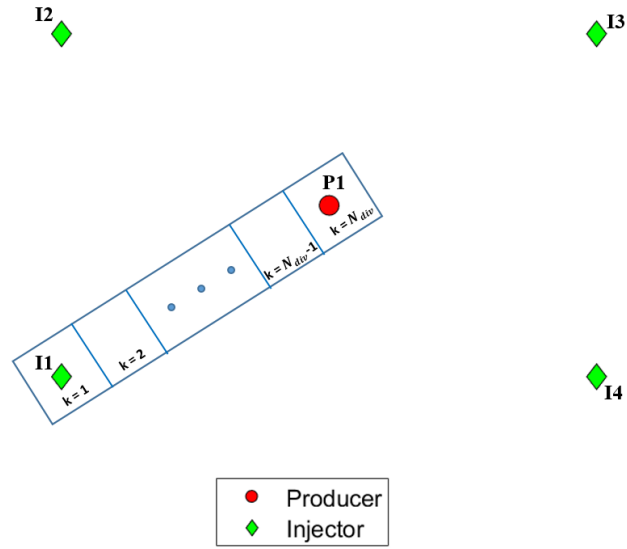

**Fig. 3** One dimensional representation of the control volume connecting Injector 1 and Producer 1.

In Eq. 14, the pore volume $\left(V_{p_{i,j}}\right)$, the dispersion coefficient $\left(D_{x_{i,j}}^{n+1}\right)$, the flow rate at block faces $\left(q_{f_{i,j}}^{n+1}\right)$, the number of divisions $\left(N_{\text{div}_{i,j}}\right)$, and the interwell connectivity factor $\left(f_{i,j}\right)$ for each control volume are unknown and should be estimated. The ISADE model therefore consists of two main steps:

1. A history matching step to esimate the values of the unknown model parameters in Eq. 14, and
2. A forward prediction step

In the history matching stage, a global optimizer is used to estimate the unknowns. A global optimizer is needed here because one of the unknown parameters (i.e., $N_{\text{div}_{i,j}}$) is a discrete variable. This makes the aplication of gradient-based estimation troublesome.

The values of the dispersion coefficient and the flow rates at the block interfaces may change with time as both parameters are affected by the injection and production rates. Therefore, to incorporate the effect of the change in well rates on these parameters, we estimate the values of the dispersion coefficient and the volumetric rate at the grid faces at a particular reference rate (taken at the first timestep) and then calculate their values at subsequent timesteps based on the reference values as

$$D_{x_{i,j}}^{n+1} = \frac{D_{x_{i,j}}^{ref}}{R_{i,j}^{ref}} \times R_{i,j}^{n+1}, \tag{15}$$

and

$$q_{f_{i,j}}^{n+1} = \frac{q_{f_{i,j}}^{ref}}{R_{i,j}^{ref}} \times R_{i,j}^{n+1}, \tag{16}$$

where, $D_{x_{i,j}}^{ref}$ and $q_{f_{i,j}}^{ref}$ are the dispersion coefficient and volumetric flow rate values at a reference rate $R_{i,j}^{ref}$, respectively. The reference rate is calculated from the injection rate $\left(I_{i,j}^1\right)$ and production rate $\left(Q_{i,j}^1\right)$ as

$$R_{i,j}^{ref} = \frac{Q_{i,j}^1 + I_{i,j}^1}{2}, \tag{17}$$

where $I_{i,j}^1$ is the injection rate from injector $i$ to producer $j$ at the first timestep and $Q_{i,j}^1$ is the production rate at the well $j$ due to injection from well $i$ also taken at the first timestep. Also, $D_{x_{i,j}}^{n+1}$ and $q_{f_{i,j}}^{n+1}$ are the dispersion coefficient and volumetric flow rate at any subsequent time-level $n + 1$, and $R_{i,j}^{n+1}$ is given by

$$R_{i,j}^{n+1} = \frac{Q_{i,j}^{n+1} + I_{i,j}^{n+1}}{2}. \tag{18}$$

Genetic Algorithm, GA (Holland, 1975) for constrained optimization was used to estimate the unknown parameters of the ISADE model. Genetic algorithm has been widely used for parameter estimation in groundwater modeling (Han et al., 2020; Kourakos & Mantoglou, 2012; McKinney & Lin, 1994). The objective function minimized in the ISADE model is

$$\Phi(\vec{\alpha}) = \sum_{j=1}^{n_p} \frac{1}{\xi_j} \left\| \vec{d}_{\text{obs}_j} - \vec{d}_{\text{cal}_j} \right\|_2, \tag{19}$$

subject to

$$\sum_{j=1}^{n_p} f_{i,j} \leq 1 \quad \forall i \in \left[1, n_{inj}\right], \tag{20}$$

where, $\vec{d}_{\text{obs}_j}$ is the vector of contaminant concentration observed at each producer $j$ (obtained either from the well history data or from full-scale simulation), $\vec{d}_{\text{cal}_j}$ is the vector of contaminant concentration observed at each producer $j$ obtained by running the ISADE model (Eq. 14), $\vec{\alpha}$ is the vector of all unknown parameters, and $\xi_j$ is the number of data points available.

Once the objective function is sufficiently minimized (i.e. a good match is found between the observed and the calculated data), the estimated parameters are then used to predict future contaminant production rates in the pumping wells.

## 4. Example Applications

Five synthetic examples are presented to test the effectiveness of the ISADE model. The first two examples considered contaminant transport in homogenous aquifers. The other examples incorporated more realistic heterogeneous aquifers with varying permeability and porosity vales. In all but Example 4, contaminants were introduced into the respective aquifers through injection wells, while in Example 4, contaminants were introduced into the aquifer from a portion of its boundary. In Example 5, a more complex aquifer with an irregular shape was considered and a commercial simulator (CMG-IMEX) was used to simulate the concentrations of contaminant at the producers. Summaries of the features of the models considered in the example are presented in Table 1. The accuracy of the ISADE model is tested by comparing the concentrations of contaminants obtained by the ISADE model to those obtained from full-scale simulation. Furthermore, the computational performance of the ISADE model is evaluated by comparing the total simulation run time of the ISADE model to that of the full-scale simulation and the. All simulations in this work were carried out on a system having an Intel(R) Xeon(R) CPU E5-1650 v3@3.50GHz processor with 16 GB RAM.

**Table 2.** Summary of Example Applications

| Example | Number of Grids in Full Scale Model | Number of Timesteps | Aquifer Heterogeneity | Number of Wells | Simulator (Full Scale Simulation) | Injection/ Production Rates | Contaminant Release Rate |
|---|---|---|---|---|---|---|---|
| 1 | 10201 | 5000 | Homogeneous | 1 Producer, 1 Injector | MATLAB Code | Constant | Constant |
| 2 | 22801 | 15000 | Homogeneous | 1 Producer, 4 Injectors | MATLAB Code | Constant | Constant |
| 3 | 22801 | 20000 | Heterogeneous | 3 Producers, 3 Injectors | MATLAB Code | Varying | Varying |
| 4 | 22801 | 20000 | Heterogeneous | 3 Producers, Injection from Aquifer Boundary | MATLAB Code | Constant | Varying |
| 5 | 38466 | 10000 | Heterogeneous | 2 Producers, 3 Injectors | CMG-IMEX | Varying | Varying |

**4.1 Example 1**

Here we consider a simple homogeneous aquifer having one injection well and one pumping well (Fig. 4). The permeability of this aquifer is 500md and the porosity is 35%. The aquifer dimensions are 5,050ft x 5,050ft x 50ft and it is divided into 101 x 101 x 1 grids. The distance between the injector and the producer is 3464.8ft. The injection and production rates vary at very early time but becomes almost constant after a short time as shown in Fig. 5. 250g/L of contaminant is continuously released into the aquifer at the injector and the pumping well produced fluids for 5000 days. The time-step size was set to 1 day. The task is to monitor the contaminant rate at the pumping well. Initially, the full-scale simulation was run to obtain the concentration of contaminants observed at the producers. The results of the ISADE model were then compared to the results of the full-scale simulation.

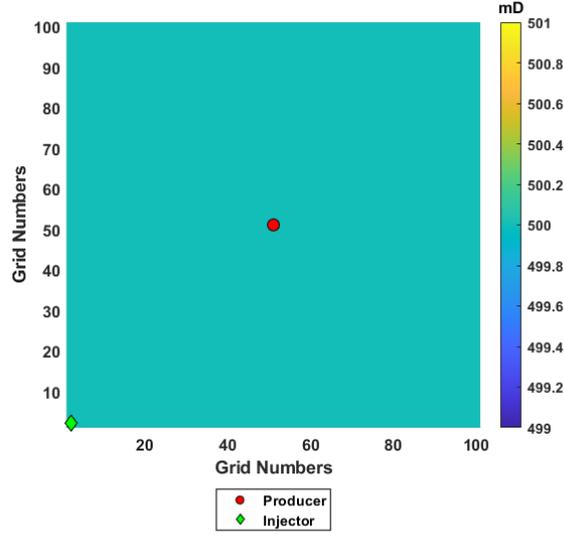

**Fig. 4** Aquifer model in Example 1 displaying the well locations.

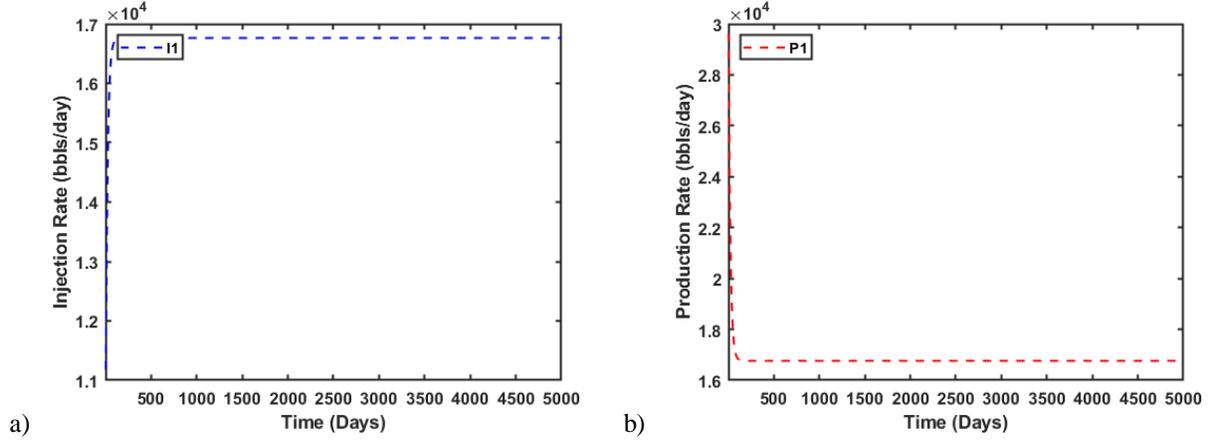

**Fig. 5** Example 1: a) injection rate and b) production rate in barrels/day.

Since, ISADE is a reduced-physics model, it requires historical production to estimate the unknown variables in order to predict future contaminant rate. In this example, contaminant rate obtained during the first half of the production time was set as the history data and used to estimate the parameters of ISADE. The contaminant rate obtained during the other half of the production time is set as the future contaminant rate and this was compared to the contaminant rate predicted by ISADE during the same time period. Table 2 lists the lower and upper bounds of the search space for the unknown variables. These limits were used in all the examples presented. Note that since the block pore volume $\left(V_{p_{i,j}}\right)$ and the the block face flow rates $\left(q_{f_{i,j}}\right)$ can potentially have very large values, these variables were estimated as exponents as

$$V_{p_{i,j}} = 10^{V^e_{p_{i,j}}}, \tag{21}$$

and

$$q_{f_{i,j}} = 10^{q^e_{f_{i,j}}}. \tag{22}$$

Table 3 displays the estimated values of different parameters obtained after the history matching process. Since this example has only one injector-producer pair it does not require the estimation of an interwell connectivity factor. Figure 6 shows the comparison between the contaminant rates obtained from the full simulation model and those obtained using ISADE. The figure shows a very good match between the results. Comparisons can also be made by calculating a mismatch value between the result obtained from ISADE and that from the full-scale simulation using the following equation

$$\Psi_j = \frac{1}{\lambda_j} \left\| \vec{c}_{ISADE_j} - \vec{c}_{FSM_j} \right\|_2, \quad (23)$$

where, $\vec{c}_{FSM_j}$ is the vector of contaminant concentration observed for each producer $j$ at each time step obtained from the full scale model, $\vec{c}_{ISADE_j}$ is the vector of contaminant concentration observed for each producer $j$ obtained by running the ISADE model, and $\lambda_j$ is the number of data values. The calculated mismatch is 0.1482 g/L thus indicating that the ISADE model was able to reliably predict the contaminant production. Table 4 shows the simulation runtime for both the full-scale simulation and ISADE. We observe from the table that the most time-consuming part of the ISADE model is in running the history matching algorithm. Table 4 also lists the speed-up factor of the ISADE model. This factor is obtained by dividing the time taken to run the full-scale simulation by the time taken to run the ISADE model. The ISADE model for Example 1 was about 23 times faster than the full-scale simulation.

**Table 2.** Search bounds for the unknown variables

| Unknown Variable | Lower Bound | Upper Bound | Unit |
|---|---|---|---|
| Interwell Connectivity Factor $(f_{i,j})$ | 0 | 1 | - |
| Exponent for Pore Volume $(V^e_{p_{i,j}})$ | 1 | 10 | - |
| Exponent for Flow rate at block faces $(q^e_{f_{i,j}})$ | 1 | 10 | - |
| Dispersion Coefficient $(D_{x_{i,j}})$ | 0.1 | 1000 | $ft^2/day$ |
| Number of Divisions $(N_{div_{i,j}})$ | 3 | 15 | - |

**Table 3.** Well distances and the estimated values for each unknown variable (Example 1)

| Well Pair | Distance (ft) | $f_{i,j}$ | $V^e_{p_{i,j}}$ | $q^e_{f_{i,j}}$ | $D_{x_{i,j}}$ | $N_{div_{i,j}}$ |
|---|---|---|---|---|---|---|
| I1-P1 | 3464.8 | - | 7.5811 | 4.8507 | 131.4904 | 4 |

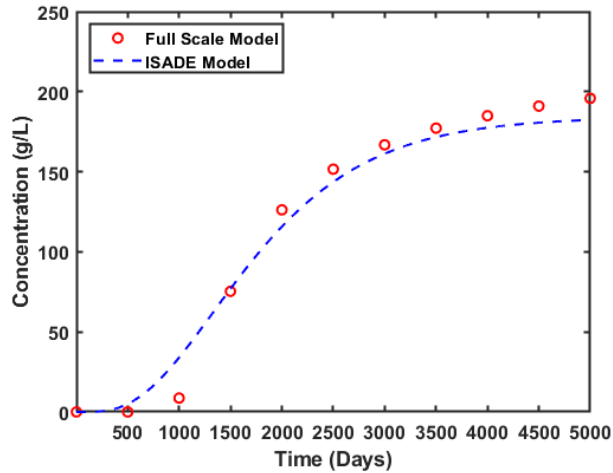

**Fig. 6** Example 1: concentration of contaminants obtained using the ISADE model and the Full Scale Model.

**Table 4.** Comparison of simulation runtimes (Example 1)

| Model | Runtime (sec.) | | Speed-up Factor |
|---|---|---|---|
| Full-scale Simulation | 270.006 | | 22.912 |
| ISADE | History matching period: 11.731 | Total: 11.784 | |
| | Prediction period: 0.053 | | |

### 4.2 Example 2

The second example is also homogeneos. The aquifer size is 5,285ft x 5,285ft x 50ft and it is divided into 151 x 151 x 1 grids. The permeability of the aquifer is set to 250md and the porosity is kept as 35%. This example, however, considers multiple injection wells. There are four injection wells and one pumping well in the aquifer (see Fig. 7). The distance between each injection and production well is shown in Table 5. The aquifer was produced for 15,000 days with a timestep of 1 day. The flowrates from the injectors and producers are displayed in Fig. 8. The first one-third of the data (data from the first 5000 days) was considered the history data was matched to obtain the unknown ISADE parameters and the remaining data (data beyond the first 5000 days) was taken as the future contaminant rate to be predicted by the ISADE model.

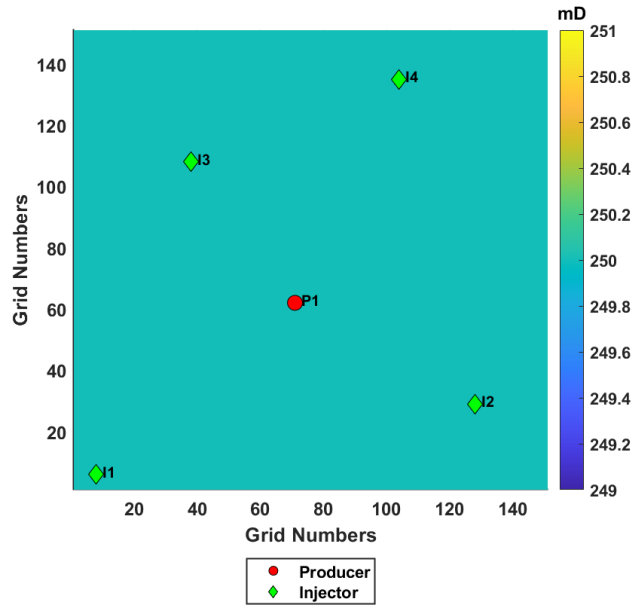

**Fig. 7** Aquifer model in Example 2 displaying the well locations.

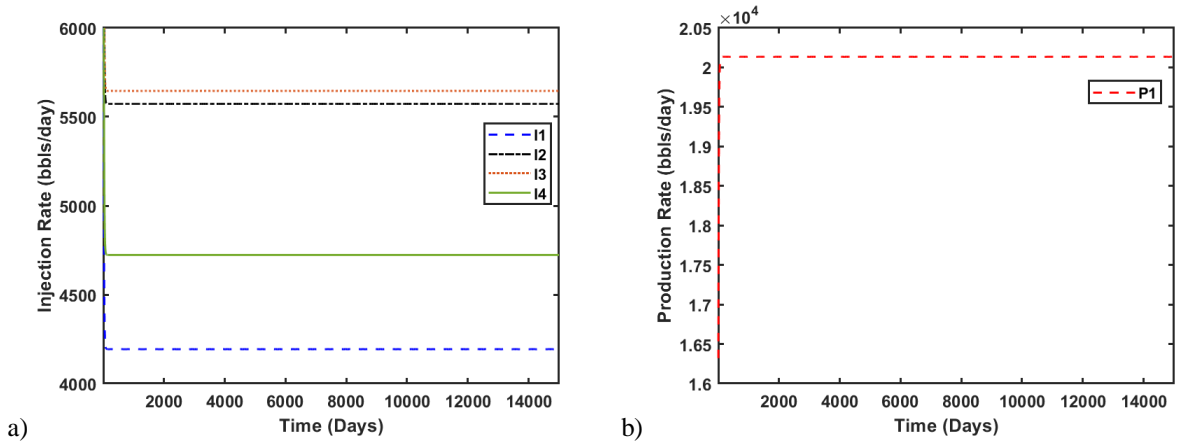

**Fig. 8** Example 2: a) injection rates and b) production rate

In this example, all five unknown ISADE parameters were to be estimated for each injector-producer pair. Thus, there are a total of 20 unknown parameters in this problem. Since there is only one producer in the aquifer we do not need to expicitly define the constraint given by Eq. 16 as it is already fulfilled by the upper limits of the interwell connectivity factors (Table 2). Table 5 shows the estimated values for each of the unknown variables. We observe that the estimated values of interwell connectivity factor for some of the wells are much less than 1. This simply indicates that not all the contaminant released at the corresponding injectors may be reaching the producer and some may remain in the aquifer. Figure 9 displays the comparison between the observed contaminant concentration obtained from full-scale simulation and that from ISADE. The figure shows a very good match between the two. The mismatch value was calculated to be equal to 0.0262 g/L. Table 6 displays the simulation runtimes for both the full-scale simulation and ISADE models. In this example, ISADE was about 10.5 times faster than the full-scale simulation.

**Table 5.** Well distances and the estimated values for each unknown variable (Example 2)

| Well Pair | Distance (ft) | $f_{i,j}$ | $V^e_{p_{i,j}}$ | $q^e_{f_{i,j}}$ | $D_{x_{i,j}}$ | $N_{\text{div}_{i,j}}$ |
|---|---|---|---|---|---|---|
| I1-P1 | 2950.2 | 0.8931 | 8.3712 | 2.9644 | 9.6822 | 4 |
| I2-P1 | 2305.2 | 0.3746 | 5.5557 | 2.7585 | 18.9962 | 5 |
| I3-P1 | 1981.4 | 0.7089 | 6.9639 | 2.7505 | 10.9077 | 5 |
| I4-P1 | 2803.9 | 0.1037 | 5.8720 | 2.6901 | 17.5790 | 4 |

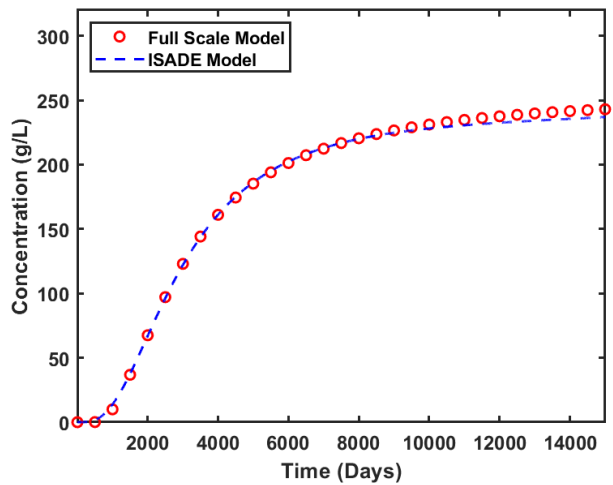

**Fig. 9** Example 2: concentration of contaminants obtained using the ISADE model and the Full Scale Model.

**Table 6.** Comparison of simulation runtimes (Example 2)

| Model | Runtime (sec.) | | Speed-up Factor |
|---|---|---|---|
| Full Scale Simulation | 2038.936 | | 10.458 |
| ISADE | History matching period: 194.695 | Total: 194.957 | |
| | Prediction period: 0.262 | | |

### 4.3 Example 3

This example consited of a heterogenous aquifer with nonuniform permeabilities and porosites as shown in Fig. 10. The aquifer dimensions are same as Example 2. There are three pumping wells and three injectors in the aquifer. The distance between each injector-producer pair is listed in Table 10. The injection and production rates are displayed in Fig. 11. The aquifer was produced for 20,000 days. One major difference in this example is that the concentration of the contaminant released into the aquifer and the well production and injection rates varies with time and is not held constant as in the previous examples (Fig. 11 & 12). Moreover, only a quarter of the data (upto 5,000 days) is used as the history data to estimate the unknown variables. Also, there are nine injector-producer pairs so that a total of 45 variables are required to be estimated in the ISADE model.

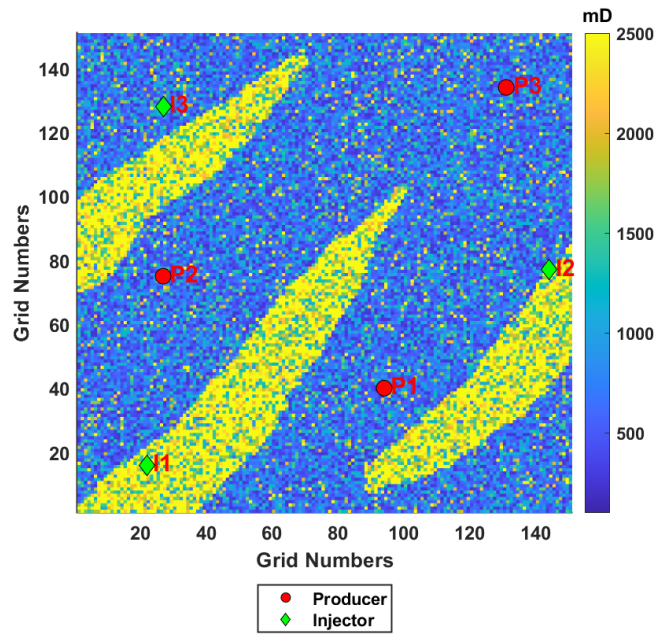

**Fig. 10** Aquifer model used in Example 3 displaying the well locations.

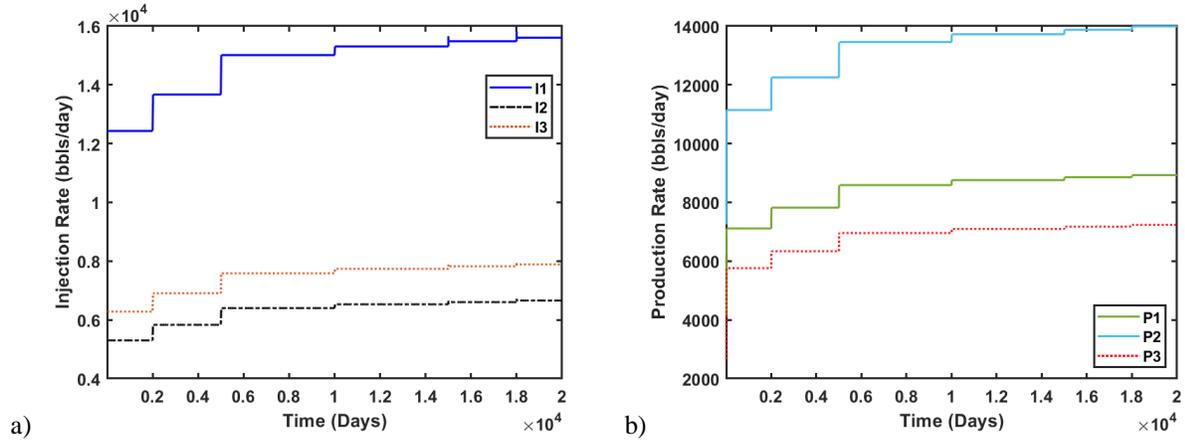

**Fig. 11** Example 3: a) injection rates and b) production rate.

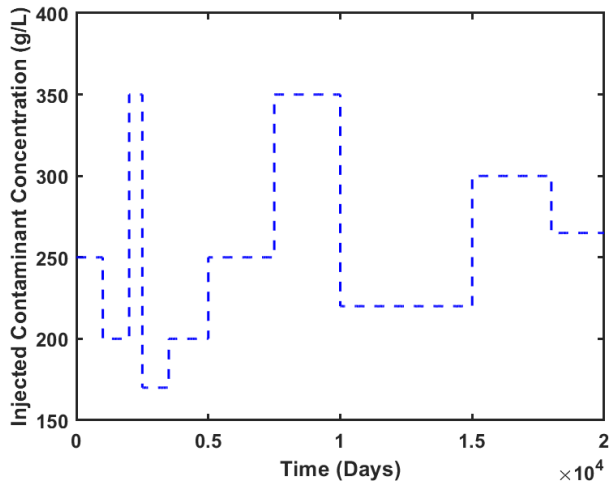

**Fig. 12** Concentration of contaminants released into the aquifer for each injector with respect to time (Example 3).

Table 7 lists the values of the estimated variables. Figure 13 displays the comparison of the contaminant concentration obtained at all the producers using the ISADE model and the full simulation model. The mismatch values for each of the producers are listed on Table 8. The data presented on both Fig. 13 and Table 8 shows that the results from ISADE are reasonably close to those from the full simulation model. The values of mismatch in this case are slightly higher than those seen in the previous examples. This is expected due to the complexities in this problem and moreover only a quarter of the available data was used for history matching. Table 9 presents a comparison of the simulation runtimes for both the full-scale simulation and ISADE. The ISADE model was about 7.3 times faster than the full-scale model.

**Table 7.** Well distances and the estimated values for each unknown variable (Example 3)

| Well Pair | Distance (ft) | $f_{i,j}$ | $V^e_{p_{i,j}}$ | $q^e_{f_{i,j}}$ | $D_{x_{i,j}}$ | $N_{\text{div}_{i,j}}$ |
|---|---|---|---|---|---|---|
| I1-P1 | 2656.3 | 0.0637 | 5.1571 | 2.1505 | 3.2634 | 4 |
| I1-P2 | 2072.4 | 0.1176 | 5.5198 | 2.4547 | 4.1554 | 3 |
| I1-P3 | 5622.4 | 0.0947 | 5.9817 | 2.2141 | 5.4494 | 3 |
| I2-P1 | 2177.0 | 0.1947 | 5.8295 | 2.2474 | 15.3947 | 3 |
| I2-P2 | 4095.6 | 0.0919 | 5.1565 | 2.2603 | 5.4630 | 4 |
| I2-P3 | 2046.2 | 0.1013 | 5.7731 | 2.1191 | 20.0095 | 3 |
| I3-P1 | 3871.1 | 0.1436 | 5.1539 | 2.1532 | 4.9358 | 4 |
| I3-P2 | 1855.0 | 0.2821 | 5.9168 | 3.0591 | 6.9888 | 4 |
| I3-P3 | 3646.1 | 0.0889 | 5.6130 | 2.2177 | 2.7468 | 3 |

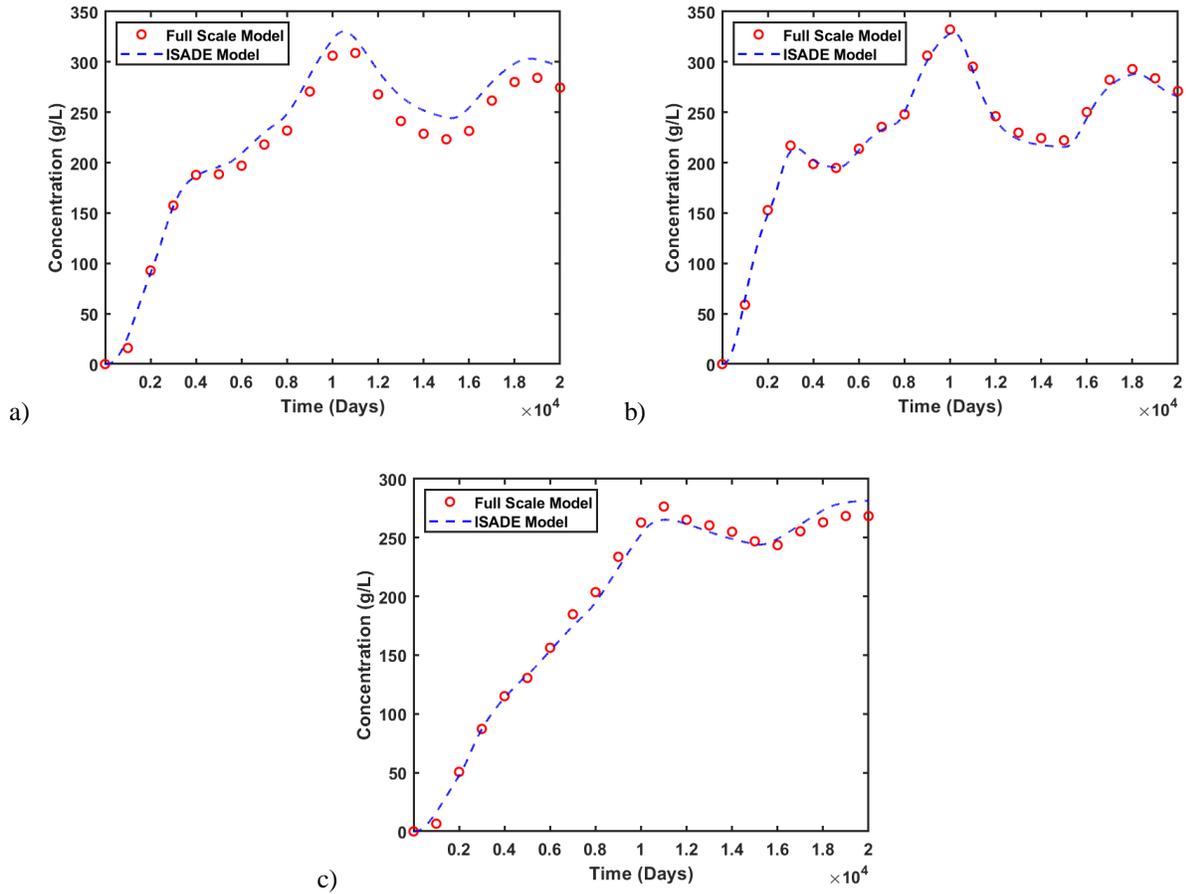

**Fig. 13** Concentration of contaminants obtained for Example 3 using the ISADE model and the Full Scale Model for a) P1, b) P2 and c) P3.

**Table 8.** Mismatch Values for each of the producers (Example 3)

| Producer | Mismatch Value (g/L) |
|---|---|
| P1 | 0.1159 |
| P2 | 0.0371 |
| P3 | 0.0516 |

**Table 9.** Comparison of simulation runtimes (Example 3)

| Model | Runtime (sec.) | | Speed-up Factor |
|---|---|---|---|
| Full Scale Simulation | 2735.156 | | 7.272 |
| ISADE | History matching period: 375.562 | Total: 376.143 | |
| | Prediction period: 0.581 | | |

**4.4 Example 4**

This example cosiders a heterogeneous aquifer which is being flooded by contaminanted water at a rate of 35000 bbls/day from the bottom left corner (Fig. 14). There are four pumping wells located in the aquifer where the concentration of the contaminant is observed. The production rate is constant and is shown in Fig. 15. The concentration of contaminant entering the system varies with time as shown in Fig. 16. The aquifer was produced for 20,000 days. Only a quarter of the data was used as history data to estimate the unknown ISADE parameters. Since there are four producers in this example, the optimizer needs to estimate 20 unknown variables. The distance between each injection-producer pair is provided in Table 10. In this example, there are no injectors, but the boundary-location and the rate at which the contaminated water enters the aquifer are known and therefore, in the ISADE model the distance between an injector-producer pair $\left(L_{well_{i,j}}\right)$ is calculated as the length between the injection location (*i*) and the corresponding producer *j* (any of P1 to P4). Moreover, the contamination through the aquifer boundary is approximated as an injection well in the ISADE model and the contamination rate within each *i-j* control volume is calculated using Eq. 9. Thus, the total injection rate $\left(I_i^{n+1}\right)$ (in Eq. 9) is assumed to be equal to the rate at which the contaminated water enters the system (35000 bbls/day).

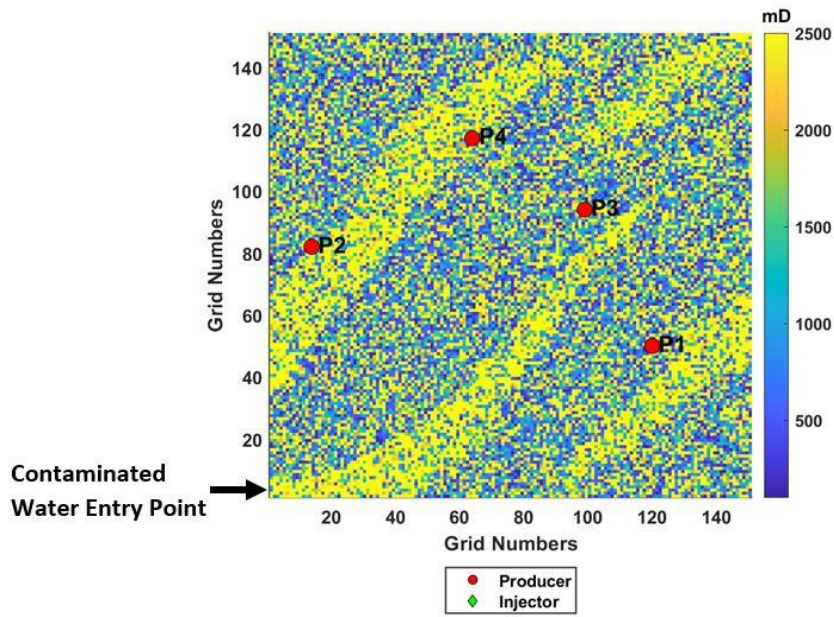

**Fig. 14** Aquifer model in Example 4 displaying the well locations.

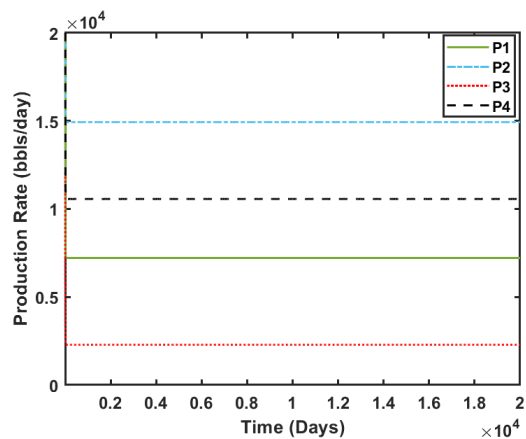

**Fig. 15** Production Rates (Example 4).

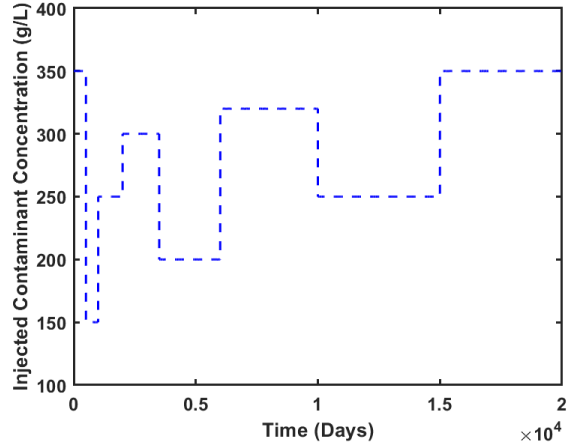

**Fig. 16** Concentration of contaminants released into the aquifer for each injector with respect to time (Example 4).

The values of the estimated variables are listed in Table 10. Figure 17 shows the comparison of the contaminant concentration for each producer obtained using the ISADE model and the full-scale simulation. The mismatch values for each producer are listed in Table 11. Results show that ISADE was able to predict the contaminant concentration in this more complex problem with reasonable accuracy. Table 12 lists the comparison of the simulation runtimes for both the full-scale simulation and the ISADE models. The ISADE model was about 13.3 times faster than the full scale model.

**Table 10.** Well distances and the estimated values for each unknown variable (Example 4)

| Well Pair | Distance (ft) | $f_{i,j}$ | $V^e_{p_{i,j}}$ | $q^e_{f_{i,j}}$ | $D_{x_{i,j}}$ | $N_{\mathrm{div}_{i,j}}$ |
|---|---|---|---|---|---|---|
| I-P1 | 2285.3 | 0.1934 | 4.2951 | 2.3736 | 21.6722 | 5 |
| I-P2 | 2750.5 | 0.4255 | 4.7714 | 3.3100 | 32.1488 | 5 |
| I-P3 | 4486.9 | 0.0683 | 3.8317 | 1.8857 | 9.6958 | 5 |
| I-P4 | 2061.6 | 0.2561 | 3.9464 | 1.9939 | 31.7861 | 5 |

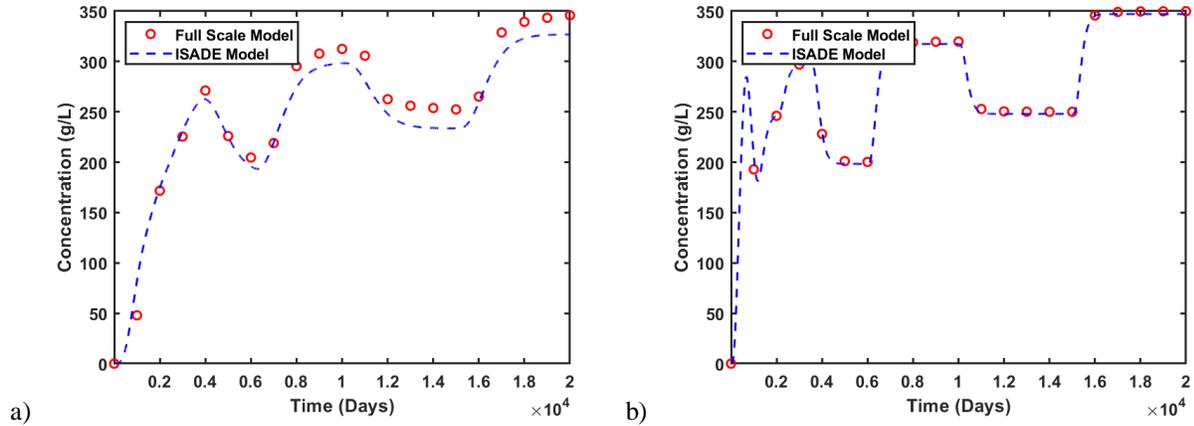

a) b)

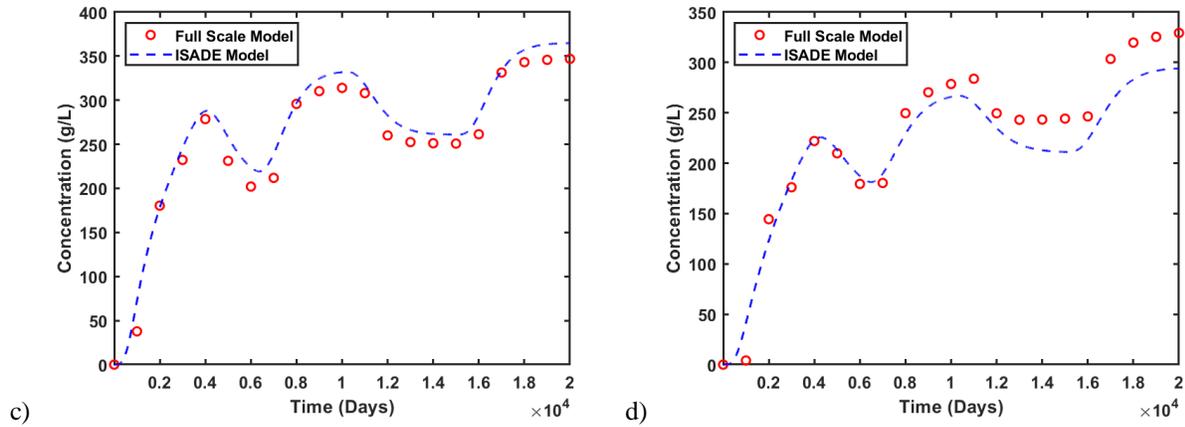

**Fig. 17** Concentration of contaminants obtained for Example 4 using the ISADE model and the Full Scale Model for a) P1, b) P2, c) P3, and d) P4.

**Table 11.** Mismatch Values for each of the producers (Example 5)

| Producer | Mismatch Value (g/L) |
|---|---|
| P1 | 0.1181 |
| P2 | 0.0453 |
| P3 | 0.1202 |
| P4 | 0.1654 |

**Table 12.** Comparison of simulation runtimes (Example 5)

| Model | Runtime (sec.) | | Efficiency of ISADE model |
|---|---|---|---|
| Full Scale Simulation | 2745.326 | | 13.277 |
| ISADE | History matching period: 206.432 | Total: 206.777 | |
| | Prediction period: 0.345 | | |

### 4.5. Example 5

In this sample application, we considered a more realistic irregular-shaped three-dimensional (gridded in $x$, $y$, and $z$-directions) heterogeneous aquifer model (Fig 18). The aquifer model is discretized into 38466 grids. The geological structure of the aquifer is based on the UNISIM-I model (Avansi et al., 2016; Avansi & Schiozer, 2015). The well rates (Fig. 19) and the concentration of the contmainants released into the system (Fig 20) vary with time. The aquifer was produced for 10,000 days with a timestep of 1 day. Only a third of the data was used as history data to estimate the unknown ISADE parameters. There were three injectors and two producers in this example and therefore the optimizer needs to estimate 30 unknown variables. The distance between each injector-producer pair is provided in Table 13.

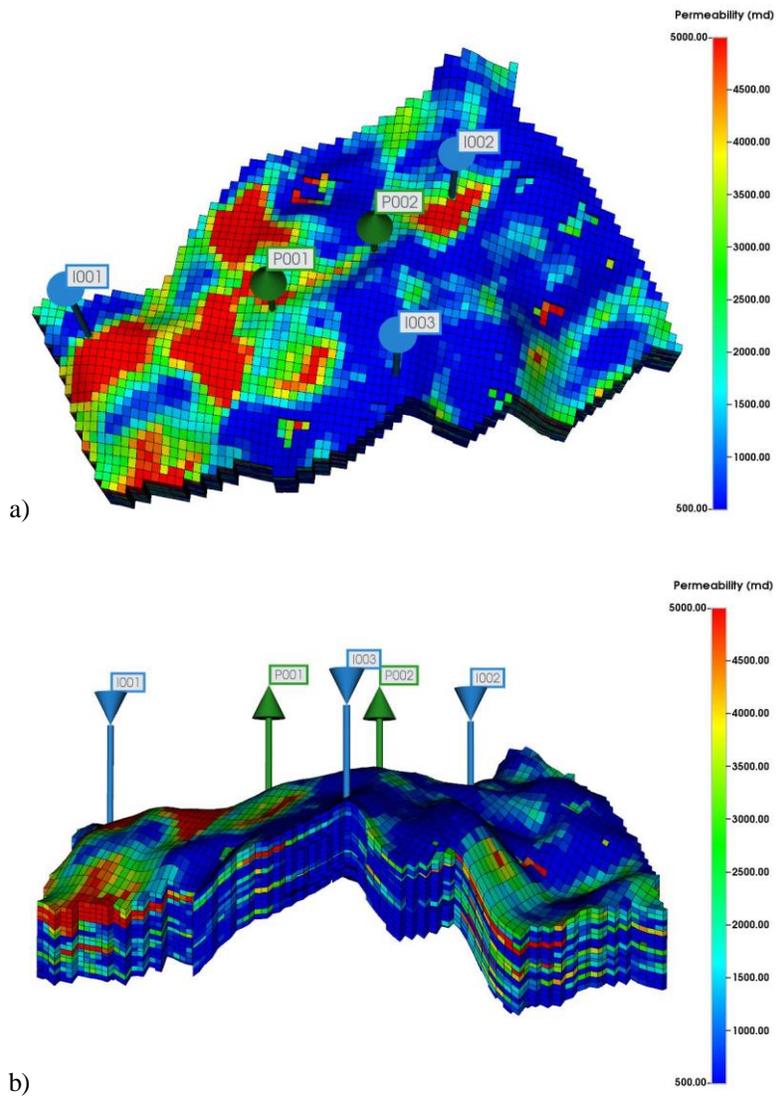

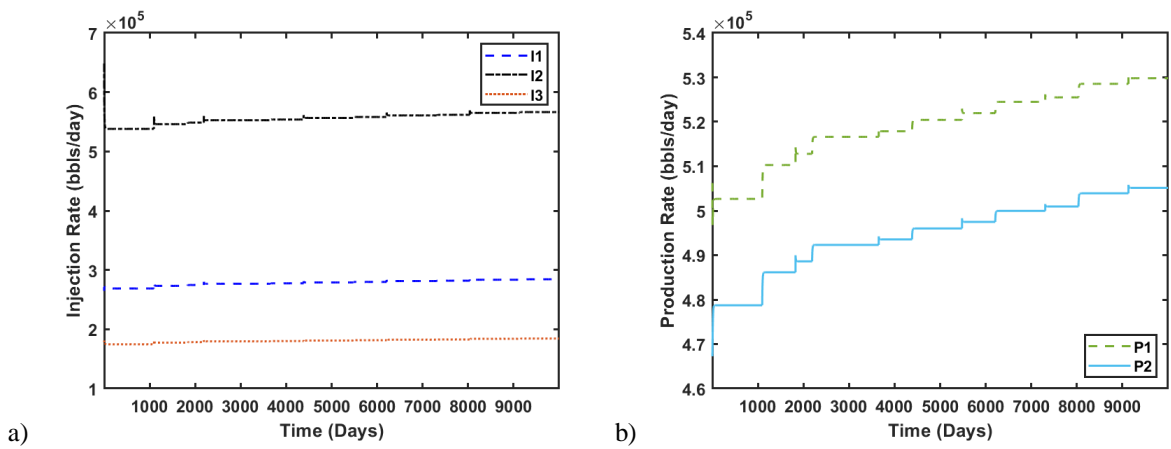

**Fig. 18** Aquifer model in Example 6 displaying the well locations for a) top view, and b) side view.

**Fig. 19** Example 6: a) injection rates and b) production rate.

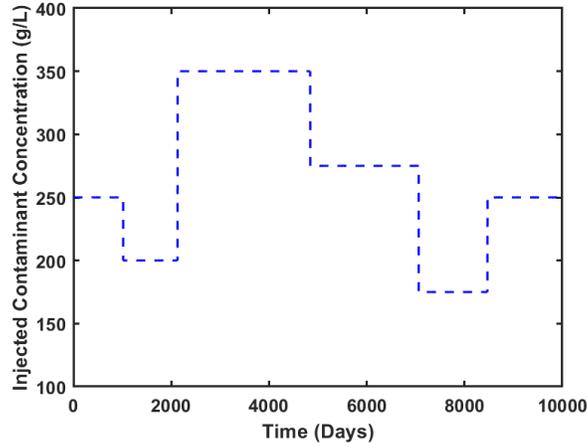

**Fig. 20** Concentration of contaminants released into the aquifer for each injector with respect to time (Example 6).

The values of the estimated variables are listed in Table 13. Figure 21 shows the comparison of the contaminant concentrations obtained using the ISADE model to those from the full-scale simulation. The mismatch values for each producer are listed in Table 14. Results show that ISADE was able to predict the contaminant concentration in this more complex problem with reasonable accuracy. Table 15 shows the comparison between the simulation runtimes of ISADE and the full-scale simulation. The ISADE model was about 14 times faster than the full-scale simulation.

**Table 13.** Well distances and the estimated values for each unknown variable (Example 5)

| Well Pair | Distance (ft) | $f_{i,j}$ | $V^e_{p_{i,j}}$ | $q^e_{f_{i,j}}$ | $D_{x_{i,j}}$ | $N_{\text{div}_{i,j}}$ |
|---|---|---|---|---|---|---|
| I1-P1 | 7446.8 | 0.1432 | 6.1960 | 2.5184 | 16.7773 | 3 |
| I1-P2 | 11832.7 | 0.1230 | 5.6390 | 3.6197 | 16.1733 | 4 |
| I2-P1 | 8570.0 | 0.1799 | 5.3393 | 2.5868 | 7.1886 | 4 |
| I2-P2 | 3961.4 | 0.2495 | 5.2078 | 2.6688 | 9.3995 | 3 |
| I3-P1 | 5476.7 | 0.2397 | 7.1515 | 4.6763 | 5.8928 | 4 |
| I3-P2 | 4840.8 | 0.2182 | 6.3955 | 2.2778 | 9.6510 | 3 |

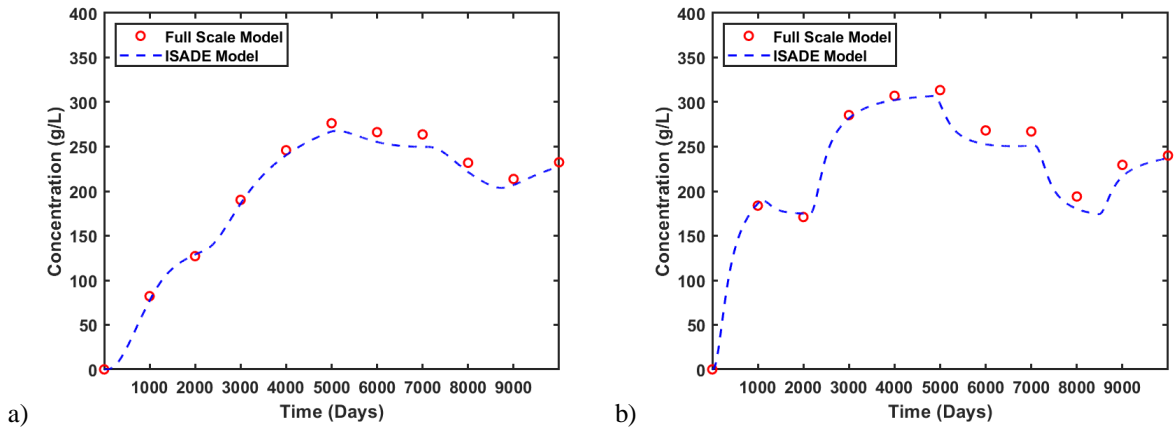

**Fig. 21** Concentration of contaminants obtained for Example 6 using the ISADE model and the Full Scale Model for a) P1 and b) P2.

**Table 14.** Mismatch Values for each of the producers (Example 5)

| Producer | Mismatch Value (g/L) |
|---|---|
| P1 | 0.0903 |
| P2 | 0.1038 |

Table 15. Comparison of simulation runtimes (Example 6)

| Model | Runtime (sec.) | | Speed-up factor |
|---|---|---|---|
| Full Scale Simulation | 2449.610 | | 14.325 |
| ISADE | History matching period: 170.791 | Total: 171.006 | |
| | Prediction period: 0.215 | | |

## 5. Discussion

The ISADE model attempts to predict contaminant concentrations observed at any producer by dividing the aquifer into a series of one dimensional injector-producer pairs. This allows for the simulation to proceed without the requirement of characterizing the properties of the whole aquifer. Five examples were presented to test the accuracy and efficiency of the ISADE model. Results indicated that the ISADE model was able to predict the concentration of contaminants at the producers even though two and three dimensional aquifers were simplified into a series of one dimensional injector-producer pairs. The success of this model can be attributed to the fact that the majority of contaminant transport occurs near the paths connecting the well pairs and therefore this simplification is able to reasonably predict the contaminant concentration observed at the production wells. During groundwater extraction, there is a push and pull occurring between every injector-producer pair. The interwell connectivity factor quantifies this phenomenon by defining the fraction of injected contaminants which is directed from a particular injector towards a particular producer. Moreover, the interwell connectivity factor also quantifies the fraction of injected contaminants that remains behind inside the aquifer and modulate the contaminant concentrations near the paths connecting the injectors and producers. This is evident by the fact that the sum of interconnectivity factors for each injector over all the producers is less than one in all the examples presented.

There are however a few limitations in the ISADE model. One limitation is that the model can only be implemented after breakthrough of contaminants because setting the parameters of ISADE requires the availability of histroy data. Secondly, ISADE assumes the values of volumetric flowrates at the grid faces and diffusion coefficients to be spatially constant. Separate values can be obtained but it would add to the computational expense of the problem.

## 6. Conclusion

In an effort to efficiently predict contaminant transport in aquifers and reservoirs we have presented the ISADE model. ISADE is a reduced-physcis model that divides the aquifer into smaller control volumes consisting of unique injector-producer pairs and then solves the one-dimensionally ADE in each control volume numerically. Interwell connectivity factors are defined to couple the several control volumes in the ISADE model. Production history data or data from full-simulation model is then used to constrain the ISADE model. ISADE has five unknown variables for each control volume that must be estimated from history matching before the model can be subsequently used for prediction. While full-scale simulation of the ADE requires the knowledge of velocities of the fluid at all grid interfaces and therefore requires the solution of the full-scale fluid flow model, ISADE only requires the production and injection rates of the wells which are either known or can be obtained using any of the fluid flow proxy models available in literature. This further reduces the total computational run time.

Five examples were presented to test the efficacy of the ISADE model. These examples showed that ISADE was able to closely match the results obtained from full-scale simulation.

**Acknowledgments**
The authors would like to acknowledge the College of Petroleum Engineering & Geosciences, King Fahd University of Petroleum & Minerals for providing the funding for this research through grant SF 20006.